\documentclass[preprint,floats,aps,showpacs]{revtex4}
\usepackage{amssymb}

\usepackage{epsfig}

\newcommand{\be}{\begin{equation}}
\newcommand{\ee}{\end{equation}}
\newcommand{\bea}{\begin{eqnarray}}
\newcommand{\eea}{\end{eqnarray}}

\newcommand{\p}{\partial}
\newcommand{\s}{\sigma}

\newcommand{\la}{\langle}
\newcommand{\ra}{\rangle}
\newcommand{\rd}{\mbox{d}}
\newcommand{\ri}{\mbox{i}}
\newcommand{\re}{\mbox{e}}

\begin{document}
\title{A Field Theory for Fermionic Ladder with Generic Intrachain Interactions.}

\author{ A. M. Tsvelik}
\affiliation{ Department of Condensed Matter Physics and Materials Science, Brookhaven National Laboratory, Upton, New York 11973-5000, USA.
}

\date{\today}

\begin{abstract}
 An effective low energy field theory is developed for a system of two chains. The main novelty of the approach is that it allows to treat  generic intrachain repulsive interactions of arbitrary strength.  The chains are coupled by a direct tunneling and four-fermion  interactions. At low energies the individual chains are described as   Luttinger liquids with an arbitrary ratio of spin $v_s$ and charge $v_c$ velocities. A judicious choice of the basis for the decoupled chains greatly  simplifies the description and allows one to separate high and low energy degrees of freedom. In a direct analogy to the bulk cuprates the resulting effective field theory distinguishes  between three qualitatively different regimes: (i) small doping ($v_c << v_s$), (ii) optimal doping ($v_s \approx v_c$) and (iii) large doping ($v_s << v_c$). I discuss the excitation spectrum and derive expressions for the electron spectral function which turns out to be highly incoherent. The degree of incoherence increases when one considers an array of ladders (stripe phase).  

\end{abstract}

\maketitle

\sloppy
\section{Introduction}

 In their seminal paper Dagotto and Rice \cite{rice} came with an idea that   one can get an insight into physics of copper oxide materials by studying just two coupled CuO chains. 
 This gave rise to the belief that the two-chain problem contains in itself seeds of the rich physics of the cuprates. Since then  the interest to the problem of ladders (that is how such two-chain systems are usually called) has never faded and there have been numerous attempts to tackle it using both numerical and analytical approaches. There are also experimental examples of fermionic  ladder materials, such as "telephone number" compound Sr$_{14-x}$Ca$_x$Cu$_{24}$O$_{41}$,  which display similarities and differences with the uniform cuprates \cite{telephone1},\cite{telephone2}.  There is a possibility that a model of weakly coupled ladders may describe the cuprates with stripe ordering (for a review, see \cite{stripes}).
 
 As far as the theory of fermionic ladders is concerned, the efforts to tackle it can be separated into numerical and analytic ones. Numerical techniques allow to  deal directly with the experimentally relevant range of the model parameters. A fermionic  ladder is usually modeled by  the $t-J$ model with equal tunneling matrix elements along the rungs and legs of the ladder, the ratio of the exchange to the tunneling matrix element is taken as $J/t = 1/3-1/2$, the ring exchange is also considered. Examples of the state-of-the-art direct diagonalization results and DMRG calculations can be found  in \cite{did1},\cite{did2} and \cite{white},\cite{marston} respectively . The use of the $t-J$ model  assumes that the on site Coulomb interaction on the individual chain is significantly greater than the tunneling matrix elements. At such circumstances one may expect a significant difference between the spin and the charge velocities for an individual chain and the numerical approaches obviously take this feature into account. The analytical approaches follow the seminal papers \cite{fisher} and are based on bosonization and renormalization group theory (see \cite{CT} and references therein). They all  start from the  limit of weak interactions and have a great problem with tackling the difference in velocities. For approximately equal spin and charge velocities there is a  remarkable result first obtained in \cite{fisher}. It turns out  that for a quite generic choice of bare coupling constants the theory scales to a strong coupling state with an enlarged U(1)$\times$O(6) symmetry greater than the symmetry of the lattice Hamiltonian. This feature allows one to solve the low energy model exactly \cite{ludwig} and to calculate its correlation functions \cite{ludwig},\cite{fabkon}. The phase diagram of the strongly correlated model contains a phase with a d-wave superconducting quasi long range order; this  has been taken as evidence that such model may be relevant for the physics of cuprates. 
 
 Both numerical and analytical approaches have their drawbacks. The present day numerical techniques used to study 2-leg ladders are exact diagonalization (Lanczos method) and DMRG. Although Lanczos method has generated many impressive results for the problem of doped ladders (I discuss them further in the text), it  does not  allow to study long chains nor (for a reasonable doping fraction)  the doping dependence. For instance, in \cite{did1} the maximal length of the system studied was 16 and the doping was represented by just two holes. There are significant size effects and it is also difficult to extract directly information about correlation functions. DMRG approach allows to work with much longer ladders (up to 100 cites in length), but to extract frequency dependent information is difficult. In any case, an analysis of numerical results frequently requires some input from field theory (see, for instance, \cite{did2},\cite{white},\cite{marston}), and it would be highly desirable to have a description valid for unequal velocities. This is just what the available analytical techniques fail to provide. Although  they  have been very successful in describing the low energy sector of the model with equal velocities, it is not clear what happens if the velocities are different, as one may expect it to be in experimentally relevant situations. Another major difficulty is related to the richness of the fermionic ladder phase diagram which includes various phases with superconducting as well as Charge Density Wave quasi-long-range order (see \cite{CT} and references therein). Since the analytic approaches start with unrealistic values of the bare interactions, parameters of the effective theory should be considered as phenomenological and one cannot, for example, trace their doping dependence.

  In this paper I present  a procedure based on bosonization and subsequent refermionization of the 2-leg ladder Hamiltonian which allows one to derive an effective low energy field theory for a generic  ratio of spin and charge velocities. My only assumption is that the resulting spectral gaps are small in comparison with the spin excitation bandwidth $\sim J$ which seems to be consistent  with the numerics \cite{did1}. The novelty of the approach lies in the judicious choice of the basis for the decoupled chains. Such choice greatly simplifies the form of the interaction and allows one to separate high and low energy degrees of freedom. Whenever is possible I compare my results with numerical ones.
  
  The paper is organized as follows. In Section II I discuss the model of doped ladder and derive the corresponding low energy effective field theory.  In Section III I discuss different regimes ( I call them underdoped, optimally doped and overdoped) with an emphasis on the underdoped one. In Section IV I discuss the problem of coupled ladders. There are separate sections for Conclusions and Acknowledgements. The paper has several appendices.
 
\section{From the lattice model to the low energy field theory}

 As a  starting point for our consideration I take the following  model of coupled chains:
 \bea
 && H = H_1 + H_2 - \sum_n t_{\perp}\Big[c^+_{n1,\s}c_{n2,\s} +h.c.\Big] + V_{12}\label{mod1}\\
 && H_i = \sum_{n}\Big\{ -t\Big[c^+_{ni,\s}c_{ni,\s} + h.c.\Big] + U N_{i\uparrow}N_{i\downarrow}\Big\} + \frac{1}{2}\sum_{n,m}V_{nm}N_{in}N_{im} \label{single}\\
 && V_{12} = \frac{1}{2}\sum_{n,m}\Big[ V_{nm,\perp}N_{1n}N_{2m} + J_{nm}{\bf S}_{1n}{\bf S}_{2m}\Big]
 \eea
 where $N = \sum_{\s}c^+_{\s}c_{\s}, ~~  S^a= \frac{1}{2}c^+_{\s}\s^a_{\s\s'}c_{\s'}$. In what follows I will assume that the interactions on chains are pretty much arbitrary and predominantly repulsive and the interactions between the chains (including the interchain tunneling $t_{\perp}$) are small in comparison to the characteristic single chain energy scales. Below I will elaborate on these restrictions. The Hamiltonian (\ref{mod1}) has U(1)$\times$SU(2)$\times$Z$_2$ symmetry.
 
 \subsection{Derivation of the low energy field theory}
 
 Away from half filling  the low energy dynamics of the single chain problem is universal and  is described by a sum of two Tomonaga-Luttinger models describing the charge and the spin sector. All details of the original lattice Hamiltonian (\ref{single}) are encoded in  few  parameters  such as the Luttinger parameters $K_c, K_s$ and spin and charge velocities $v_s,v_c$.  The SU(2) symmetry of the spin sector fixes the value of $K_s$ to be one. In the future we set $K_s=1$ and drop the subscript for $K_c$ so that $K_c = K$. 
    The   Lagrangian of two noninteracting chains is 
  \bea
  && L_0 = \int \rd x\Big[({\cal L}_{1c} + {\cal L}_{2c}) + ({\cal L}_{1s} + {\cal L}_{2s})\Big]\\
  &&{\cal L}_{ic} = \frac{1}{2K}\Big[v_c^{-1}(\p_{\tau}\Phi_{ic})^2 + v_c(\p_x\Phi_{ic})^2\Big], ~~ i =1,2\label{charge}\\
  && {\cal L}_{is} = \frac{1}{2}\Big[v_s^{-1}(\p_{\tau}\Phi_{is})^2 + v_c(\p_x\Phi_{is})^2\Big] - 2\pi v_s g {\bf J}_R{\bf J}_L,\label{spin}
\eea
where $\Phi_{ic},\Phi_{is}$ are bosonic fields from the charge ($c$) and spin ($s$ sectors) of $i$-th chain. The last term in (\ref{spin}) represents a marginal  interaction of right- and left-moving spin currents $J_{R,L}^a$. These currents satisfy the SU$_1$(2) Kac-Moody algebra;  their explicit expressions  in terms of bosonic fields are not necessary for the present discussion. For systems with predominantly repulsive interactions $g > 0, g \sim 1$ the current-current interaction is marginally irrelevant. The present universal description is valid  below certain doping dependent  cut-off $\Lambda(\delta)$ ($\delta$ is doping). For the Hubbard model with strong on site repulsion $U >> t$ the cut-off is 
\be
\Lambda = \mbox{min}\left[J \approx 4t^2/U,4t\sin^2(\pi\delta/2)\right].
\ee
The ratio $v_s/v_c$ is also doping dependent since $v_c$ as a function of doping has a maximum vanishing  in the limit of zero doping  and $v_s$ is weakly doping dependent (see, for example, \cite{essler}).

 My goal now is to choose a convenient basis of fields to treat the coupled chains. The first step is standard. Namely, I  introduce bosonic fields 
 \be
 \Phi^{(\pm)}_{c,s} = \Big[\Phi^{(1)} \pm \Phi^{(2)}\Big]_{c,s}/\sqrt 2 \label{trans}
 \ee
 together with their chiral components $\phi,\bar\phi = (\Phi \pm \Theta)/2$, $\Theta$ being the field dual to $\Phi$. Let us consider, for instance,  charge Lagrangian (\ref{charge}). After transformation (\ref{trans}) it becomes 
 \bea
 {\cal L}_{1c} + {\cal L}_{2c} = \frac{1}{2K}\sum_{a=\pm}\Big[v_c^{-1}(\p_{\tau}\Phi_{c}^{(a)})^2 + v_c(\p_x\Phi_{c}^{(a)})^2\Big].
 \eea
 Away from half filling the Umklapp processes are suppressed and field $\Phi_c^{(+)}$ decouples from the interactions (this statement is also supported by the direct calculations presented below). As far as the $(c,-)$ Lagrangian is concerned,  I refermionize it:
 \bea
 && \frac{1}{2K}\Big[v_c^{-1}(\p_{\tau}\Phi_{c}^{(-)})^2 + v_c(\p_x\Phi_{c}^{(-)})^2\Big] = \nonumber\\
 && r^+(\p_{\tau} - \ri v_c\p_x)r + l^+(\p_{\tau} + \ri v_c\p_x)l + g_K v_cr^+rl^+l, \label{cminus}
 \eea
 where 
 \bea
 r = \frac{\kappa}{\sqrt{2\pi a_0}}\re^{\ri\sqrt{4\pi}\phi_c^{(-)}}, ~~ l = \frac{\kappa}{\sqrt{2\pi a_0}}\re^{-\ri\sqrt{4\pi}\bar\phi_c^{(-)}}, ~~ \kappa^2 =1, \label{bosoniz}
 \eea
 with  $\kappa$ and $a_0$ being a coordinate independent Klein factor and  a small distance cut-off respectively. The coupling $g_K$ is related to the Luttinger parameter $K$; for $|K -1| << 1$ we have:
 \be
 g_K \approx 2\pi (1/K -1)
 \ee
 In view of the further developments I introduce Majorana fermions $\xi^3_{R,L}$ and $\eta_{R,L}$ related to the  real and imaginary parts of conventional fermions $r,l$:
 \bea
 r = (\xi^3_R + \ri \eta_R)/\sqrt 2, ~~ l = (\xi^3_L + \ri\eta_L)/\sqrt 2, ~~ {\xi^3}^+ = \xi^3, \eta^+ = \eta. \label{maj1}
 \eea
 In the bosonic language these Majorana operators are
 \bea
 \eta_R = \frac{\tilde \kappa}{\sqrt{\pi a_0}}\sin(\sqrt{4\pi}\phi_c^{(-)}), \xi^3_R = \frac{\tilde \kappa}{\sqrt{\pi a_0}}\cos(\sqrt{4\pi}\phi_c^{(-)}).
 \eea
 Majorana fermion fields  can be expanded in terms of conventional creation and annihilation operators. For instance, we have
  \bea
  \eta(x) = \sum_{k >0}[\eta_k \re^{-\ri kx} + \eta_k^+\re^{\ri kx}], ~~ \{\eta_k, \eta_p^+\} = \delta_{kp}.
  \eea
Substituting (\ref{maj1}) into (\ref{cminus}) I obtain the Majorana fermion form of the Lagrangian for the antisymmetric charge mode:
\bea
&& L_{(c,-)} = \frac{\ri}{2}\eta_R(\p_{\tau} - v_c\p_x)\eta_R + \frac{\ri}{2}\eta_L(\p_{\tau} + v_c\p_x)\eta_L + \frac{\ri}{2}\xi^3_R(\p_{\tau} - v_c\p_x)\xi^3_R + \nonumber\\
&& \frac{\ri}{2}\xi^3_L(\p_{\tau}  + v_c\p_x)\xi^3_L +  2 g_K v_c\eta_R\eta_L\xi^3_R\xi^3_L. \label{cminus1}
\eea

 Similar transformations in the spin sector (see \cite{shelton} for details) yield the following Lagrangian density:
 \bea
 && L_s = \frac{1}{2}\sum_{a=0}^3\Big[\chi_R^a(\p_{\tau} - \ri v_s\p_x)\chi_R^a + \chi_L^a(\p_{\tau} + \ri v_s\p_x)\chi_L^a \Big]  + V_{ex} \label{2spin}\\
 && V_{ex} = -\pi v_s g\sum_{i>j}(\chi_R^i\chi_L^i)(\chi_R^j\chi_L^j),\label{V}
 \eea
 where $\chi_{R,L}^a$ are also Majorana fermions defined as 
 \bea
 && \chi_R^0 = \frac{\tilde \kappa_{s-}}{\sqrt{\pi a_0}}\sin(\sqrt{4\pi}\phi_s^{(-)}), ~~ \chi_R^3 = \frac{\tilde \kappa_{s-}}{\sqrt{\pi a_0}}\cos(\sqrt{4\pi}\phi_s^{(-)}),\nonumber\\
 && \chi_R^1 = \frac{\tilde \kappa_{s+}}{\sqrt{\pi a_0}}\sin(\sqrt{4\pi}\phi_s^{(+)}), ~~ \chi_R^2 = \frac{\tilde \kappa_{s+}}{\sqrt{\pi a_0}}\cos(\sqrt{4\pi}\phi_s^{(+)}).
 \eea
  with $\tilde\kappa_{s,\pm}$ being new Klein factors. In the process of derivation I used the fact that 
  \bea
  J_R^{1,a} + J_R^{2,a} = \frac{\ri}{2}\epsilon_{abc}\chi_R^b\chi_R^c, ~~ J_R^{1,a} - J_R^{2,a} = \ri\chi_R^0\chi_R^a,
  \eea
  with similar expressions for the left currents. 
 
  Now let us consider the interchain tunneling:
  \bea
  V_t = t_{\perp}\sum_n(c^+_{1n,\s}c_{2n,\s} + h.c.) \approx t_{\perp}\int \rd x (\psi^+_{1R,\s}\psi_{2R,\s} + \psi^+_{1L,\s}\psi_{2L,\s} + h. c.) +..., \label{tunn}
  \eea
  where 
  \bea
  c_n = \re^{-\ri k_F n}\psi_R(x) + \re^{\ri k_F n}\psi_L(x), ~~ x = na \label{original}
  \eea
  and the dots stand for interaction terms generated by virtual high energy processes. The latter  terms contribute to the renormalization of interachain exchange and density-density interactions. 
  Using for $\psi_{R,L}$ bosonization formulae similar to (\ref{bosoniz})  I obtain:
  \bea
  \psi^+_{1R,\s}\psi_{2R,\s} + h. c = \frac{Z^2}{2\pi a_0}\sum_{\s}\kappa_{1\s}\kappa_{2\s}\Big[\re^{\ri\sqrt{4\pi}\phi_c^{(-)}}\re^{\ri\s\sqrt{4\pi}\phi_s^{(-)}} - h.c.\Big],
  \eea
  where $Z$ is a nonuniversal amplitude which magnitude depends on the intra-chain interaction. Taking into account that the combinations of Klein factors $\kappa_{1\uparrow}\kappa_{2\uparrow}$ and $\kappa_{1\downarrow}\kappa_{2\downarrow}$ commute with each other and their squares are equal to -1, we can choose them as equal to $\pm \ri$ and get for (\ref{tunn}) 
  \bea
  V_t \approx 2\ri t_{\perp}'\int \rd x \Big[\eta_R\chi_R^0 + \eta_L\chi_L^0\Big] + \mbox{interchain exchange}, \label{Vt}
  \eea
 where $t_{\perp}' = Z^2t_{\perp}$. In what follows I'll drop the prime superscript at $t_{\perp}'$.  Recall that $\chi_0$ is one of the Majorana fermions from the spin sector. However,  since it transforms as a singlet under action of the SU(2) group, the entire tunneling term is an SU(2) singlet, as it must be. Naturally, the tunneling entangles the charge and the spin sectors, but only the singlet fermion participates. As the next step I diagonalize the quadratic part of the Largangian (\ref{cminus1}, \ref{Vt}, \ref{2spin}) containing $\eta$ and $\chi^0$.   Switching to the Hamiltonian formalism and dropping the superscript and subscript  for $\chi^0_R$ I get the following Hamiltonian  for the right moving sector:
  \bea
  H' = \sum_{k>0}\Big[v_c k\eta^+_k\eta_k + v_s k \chi^+_k\chi_k + t_{\perp}'(\chi^+_k\eta_k + \eta^+_k\chi_k)\Big],
  \eea
  with $\eta_k,{\eta_k}^+,\chi_k,\chi_k^+$ being conventional creation and annihilation operators. 
  The spectrum is
  \bea
  E_{\pm}(k) = (v_c + v_s)k/2 \pm \sqrt{(v_c-v_s)^2k^2/4 + {t_{\perp}'}^2}, ~~ k>0. 
  \eea
  Linearizing the spectrum close to the point $Q = t_{\perp}'/\sqrt{v_sv_c}$ where $E_-(Q) =0$ and ignoring the upper branch of the spectrum which lays at high energies: $E_+ > t_{\perp}'$, I get 
  \bea
  && H' \approx \sum_{|k| << Q} u k {\cal R}^+_k{\cal R}_k, ~~ u = \frac{2v_cv_s}{v_c + v_s}, \\
  && \chi_R^0(x) \approx \Big[\frac{v_c}{v_s + v_c}\Big]^{1/2}\Big(\re^{-\ri Q x}{\cal R} + \re^{\ri Q x} {\cal R}^+\Big) = \Big[\frac{2v_c}{v_s + v_c}\Big]^{1/2}\Big(\cos Qx \xi^1_R + \sin Qx \xi^2_R\Big),\label{chi}\\
  && \eta_R(x) \approx \Big[\frac{v_s}{v_s + v_c}\Big]^{1/2}\Big(\re^{-\ri Q x}{\cal R} - \re^{\ri Q x} {\cal R}^+\Big) = \Big[\frac{2v_s}{v_s + v_c}\Big]^{1/2}\Big(-\sin Qx \xi^1_R + \cos Qx \xi^2_R\Big)\label{xi}.
  \eea
  where ${\cal R}, {\cal R}^+$ is a conventional (Dirac) right-moving fermion defined on the entire $k$ axis and $\xi^{(1,2)}$ are its Majorana components (its real and imaginary parts). For the left-moving fermions I get  the same expressions with $u \rightarrow -u$ and $Q \rightarrow -Q$. Substituting (\ref{xi}) into (\ref{cminus1}) for energies less than $t_{\perp}'$ I obtain the following Hamiltonian density:
  \bea
  && H_{c,-} + H_s = H_0 + V_{int},\nonumber\\
  && H_0 =  \label{H0}\\
  && \frac{\ri v_c}{2}(-\xi^3_R \p_x \xi^3_R + \xi_L^3\p_x\xi^3_L) + \frac{\ri u}{2}\sum_{a=1,2}(-\xi^a_R\p_x \xi_R^a + \xi_L^a\p_x \xi_L^a)  +  \frac{\ri v_s}{2}\sum_{a=1}^3(-\chi^a_R\p_x \chi_R^a + \chi_L^a\p_x \chi_L^a), \nonumber
  \eea
  where $V_{int}$ contains four-fermion terms. At this point it is worth to discuss briefly the problem of Fermi points. Naturally, one expects that the interchain tunneling splits the original Fermi point in two. From the above discussion one may get an impression that there is just one Fermi point. This is certainly not the case since even from Eqs.(\ref{chi},\ref{xi}) one can see that there are oscillations at wave vectors $k_F \pm Q$. To check that no low energy modes got missing on the way, I compare the central charge of (\ref{H0}) with the central charge of theory of four noninteracting fermions. The latter one is equal to 4; central charge C=1 goes into the symmetric charge mode $(c,+)$ which decouples from the rest of the system. What remains is C=3 which coincides with the central charge of model of six Majorana fermions (\ref{H0}). Indeed, every Majorana mode representing a half of conventional fermion carries central charge 1/2 so that   $1/2\times 6 =3$. 

  To obtain the  interacting part of the Hamiltonian we have to take into account the four-fermion interaction in (\ref{cminus}), as well as (\ref{V}) and the properly bosonized and refermionized interchain interaction. The latter one has the following form (\cite{sheltsv})
  \bea
    \frac{\ri}{\pi a_0} \cos(\sqrt{4\pi}\Phi_c^{(-)})\Big[ (-J+V)\sum_{a=1}\chi_R^a \chi_L^a + (3J +V)\chi_R^0\chi_L^0\Big] + \tilde V(\p_x\Phi_c^{(-)})^2 
\eea
Here $J$ and $V$ are related to the interchain exchange and density-density interaction and $\tilde V$ to the density-density forward scattering. Since the forward scattering usually yields only weak corrections to coupling constants,  I will put $\tilde V =0$. In terms of the Majoranas we have
\bea
&& \frac{1}{2\pi a_0}\cos(\sqrt{4\pi}\Phi_c^{(-)}) = \frac{\ri}{2}(\eta_R\eta_L + \xi^3_R\xi^3_L) = \frac{\ri}{2}\Big\{\xi^3_R\xi^3_L + \frac{v_s}{(v_s + v_c)}[\xi^1_{R}\xi^1_{L} - \xi^2_{R}\xi^2_{L}]\Big\}\nonumber\\
&& \chi^0_R\chi_L^0 = \frac{v_c}{v_c + v_s}[\xi^1_{R}\xi^1_{L} - \xi^2_{R}\xi^2_{L}]
\label{cos}
\eea
As it was mentioned above, the symmetric charge mode $\Phi_c^{(+)}$ does not appear in the interactions and thus decouples. 

\begin{figure}[ht]
\begin{center}
\epsfxsize=0.65\textwidth
\epsfbox{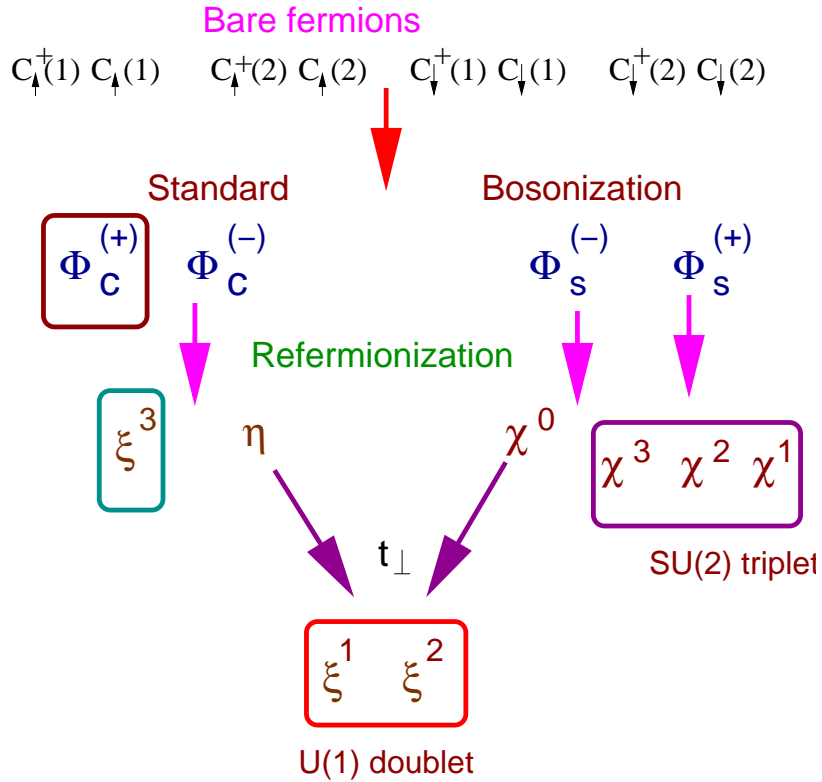}
\end{center}
\caption{The succession of transformations leading from the original formulation (1) to Eq.(\ref{H0}) . The framed fields is the ones taking part in the final effective theory. }
\label{bosonization}
\end{figure}

Combining (\ref{V}) and (\ref{cos}) I obtain
\bea
&& V_{int} = -2(\xi^3_R\xi^3_L)\Big[g_{\s,-}(\chi_R^a\chi_L^a) +  g_{c,ss}(\xi_R^1\xi_L^1 - \xi_R^2\xi_L^2)\Big] - g_{\rho,-}(\xi^1_R\xi_l^1 - \xi_R^2\xi_L^2)^2 \nonumber\\
&& - 2g_{c,st}(\xi^1_R\xi_l^1 - \xi_R^2\xi_L^2)\sum_{a=1}^3(\chi_R^a\chi_L^a) - 2g_{\s,+}\sum_{a>b,a,b=1}^3(\chi_R^a\chi_L^a)(
 \chi_R^b\chi_L^b) \label{Int}
\eea
where
\bea
&& g_{\s,-} = (V - J)/2, ~~ g_{c,ss} = v_c\Big(- g_K v_s + V + 3J\Big)/2(v_c +v_s), ~~ g_{\rho_-} =  \frac{(v_sv_c)(V +3J)}{(v_c+v_s)^2}, \nonumber\\
&& g_{c,st} = \Big[(V-J)/2 + \pi g v_c\Big]\frac{v_s}{v_c + v_s}, ~~ g_{\s,+} = 
 \pi g v_s. \label{Bare}
\eea
The complete effective Hamiltonian is a combination of the Luttinger Hamiltonian for $\Phi_c^{(+)}$ and (\ref{H0},\ref{Int}). It  has a similar  form to the one obtained  by the weak coupling  bosonization procedure (see, for instance, \cite{CT} which notations I follow). There is one important difference: the velocities of different Majorana modes are now different.  The resulting Hamiltonian, as expected, has the same Z$_{2} \times$U(1)$\times$SU(2) symmetry as the original lattice model (1). In that sense its general form is dictated by the symmetry considerations. However, the microscopic derivation is not without use since it puts constraints on the model parameters which I am  going to exploit. 

 \subsection{Some important operators.}
 
  Since the resulting field theory  is qualitatively the same as the one described in \cite{CT}, all expressions for possible order parameters (OP) are the same as in \cite{CT}. Since I am going to concentrate on conditions for superconducting pairing, only two OPs are  needed, one of them being  the  
  superconducting OP antisymmetric under chain permutation. In one dimension this property imitates  $d$-wave symmetry. Since a model of a single  Majorana fermion is equivalent to the Quantum Ising model (see Appendix A, a more detailed discussion of this subject is given, for instance, in \cite{shelton},\cite{book}), it is convenient to express the order parameters in terms of order  and disorder operators of the corresponding Ising models: $\s,\mu$ (for $\chi$ fermions) and $\Sigma, M$ (for $\xi$ ones). So we have  
  \bea
  \Delta_{SCd} \sim \re^{\ri\sqrt\pi\Theta_c^{(+)}}\Big\{ \s_1\s_2\s_3\Sigma_3\Sigma_1 M_2 - \ri M_3 M_1\Sigma_2\mu_1\mu_2\mu_3\Big\}. \label{SCdOP}
  \eea
  
   The other important operator is  the $4k_F$ component of the particle density. Its importance is related to the fact that  the $4k_F$ instability corresponding to crystallization of pairs competes with the $d$-wave superconductivity. Meanwhile, the form of this operator remains controversial. In \cite{CT} it  was conjectured on symmetry grounds, as in other publications (see, for instance, \cite{marston}). Here I provide a rigorous derivation. For the Hubbard model we have
  \bea
  \rho(4k_F,x) = F\cos(4k_F x + \sqrt{8\pi}\Phi_c)
\eea
where $F$ is a non-universal amplitude. For the fermionic  ladder one can define symmetric and antisymmetric combinations of charge densities:
\bea
&& \rho^{(+)}(4k_F,x) = 2F\cos(4k_Fx +\sqrt{4\pi}\Phi_c^{(+)})\cos(\sqrt{4\pi}\Phi_c^{(-)}) \approx \nonumber\\
&& \ri F\cos(4k_Fx +\sqrt{4\pi}\Phi_c^{(+)})\Big\{\xi^3_R\xi^3_L + \frac{v_s}{(v_s + v_c)}[\xi^1_{R}\xi^1_{L} - \xi^2_{R}\xi^2_{L}]\Big\}\label{4kF}
\eea
and
\bea
&& \rho^{(-)}(4k_F,x) = 2F\sin(4k_Fx +\sqrt{4\pi}\Phi_c^{(+)})\sin(\sqrt{4\pi}\Phi_c^{(-)}) = \nonumber\\
&& 2F\sin(4k_Fx + \sqrt{4\pi}\Phi_c^{(+)})[\eta_R\xi^3_L + \xi^3_R\eta_L]\approx \nonumber\\
&& \ri F\sqrt{v_s/(v_c + v_s)}\sin(4k_F x + \sqrt{4\pi}\Phi_c^{(+)})\Big[\re^{-\ri Q x}({\cal R}\xi^3_L + \xi^3_R {\cal L}^+) - h. c.\Big]. \label{4kF-}
\eea
The latter formula indicates that  weaker harmonics appear on wave vectors $4k_F \pm Q$. 

 The standard approach to fermionic  ladders maintains that when the system scales to strong coupling and gaps emerge in various sectors, amplitudes of the corresponding OPs freeze and only the phase factors containing gapless charge mode $\Phi_c^{(+)}$ or its dual $\Theta_c^{(+)}$ remain. As a result the scaling dimensions of the OPs become much smaller than for noninteracting fermions leading to a strong enhancement of the corresponding susceptibilities. For instance, the SCd OP (\ref{SCdOP}) being a bilinear combination of fermions; in the absence of interactions  its scaling dimension $d_{bare} = 1$, for decoupled chains $d_{decoupled} = 3/4 + 1/4K > 1$. On the other hand, in the SCd phase of 2-leg ladder  when the amplitude 
 \bea
 A = \Big\{ \s_1\s_2\s_3\Sigma_3\Sigma_1M_2 - \ri M_3 M_1\Sigma_2\mu_1\mu_2\mu_3\Big\}. \label{A}
 \eea
 acquires a nonzero ground state average the scaling dimension becomes 
 \be
 d_{SCd} = 1/4K < 1.
 \ee
 One may think that $2k_F$ Charge Density Wave OPs containing $\re^{\ri\sqrt\pi\Phi_c^{(+)}}$ exponents with scaling dimension $K/4$ are more relevant when $K< 1$ (repulsive interactions). This, however, is not the case because it turns out that in the phase with nonzero SCd amplitude (\ref{A}) the amplitudes of all other OPs (except of the $4k_F$ density one!) do not have nonzero ground state averages. The amplitudes of (\ref{4kF-}) also can never form because of 
the symmetry ($\xi_R^a$ and $\xi_L^b$ with $a\neq b$ never pair).

\section{Different regimes. }

 Although model (\ref{H0},\ref{Int}) is not integrable, some general statements can be made concerning its spectrum, especially in the regions where two of the velocities $v_c$ and $v_s$ are very different.  Since  interactions (\ref{Int}) are marginal, for general value of parameters they can all flow either to weak or strong coupling. The numerics done on two-leg ladders usually points to the strong coupling regime. However, a possibility of the weak coupling also exists \cite{piotr1},\cite{piotr2} and, as I will discuss further,  apparently is realized in this model for $v_c \sim v_s$ when one can apply the standard RG. 
 
  As I am going to argue, the strong coupling regime is realized when the velocities are very different so that one excitation branch lays well above the other in most of the phase space. 
 In that case I use the adiabatic approximation. 
 Such  approximation has been used extensively to study soliton excitations in the Peierls-Fr\"ohlich model pertaining to the polyacetylene problem where the slow subsystem is represented by the  lattice phonons. The representative papers on the model in question are \cite{braz}, the summary can be found in review articles \cite{kivelson},\cite{kirova}. 
  
\subsection{Underdoped regime $v_c << v_s$. Adiabatic approximation} 
 
  In the given case the fast particles are spin Majorana fermions $\chi_a$.  I bosonize slow fermionic modes $\xi^1,\xi^2$ so that the Hamiltonian (\ref{H0},\ref{Int}) becomes 
  \bea
 && {\cal L} = \frac{1}{2}\left[u^{-1}(\p_{\tau} \theta)^2  + u(\p_x\theta)^2\right] + \frac{1}{2}\Big[\xi^3_R(\p_{\tau} -\ri v_c \p_x) \xi^3_R + \xi_L^3(\p_{\tau} + \ri v_c\p_x)\xi^3_L\Big] \nonumber\\
 && \frac{\Delta^2}{2\gamma}  + \ri\Delta\Big[\frac{g_{s,cc}}{g_{s,st}}(\xi_R^3\xi_L^3) + \sum_{a=1}^3(\chi_R^a\chi_L^a)\Big]  + \nonumber\\
 &&  \frac{1}{2}\sum_{a=1}^3\Big[\chi^a_R(\p_{\tau} - \ri v_s\p_x) \chi_R^a + \chi_L^a(\p_{\tau} +\ri v_s\p_x) \chi_L^a\Big] + (\mbox{4-fermion interaction}),\label{PF}\\
  && \Delta = \frac{g_{s,st}}{\pi a_0}\cos[\sqrt{4\pi}\theta], ~~ \gamma = 4{g_{c,st}}^2/g_{\rho,-}. \nonumber
 \eea
  Here it is assumed that $g_{\rho,-} > 0$. In the limit $u << v_s$ field $\theta$ becomes static \cite{omega}. The $\xi^3$ fermion being nonclassical object on this stage  should  be neglected. I will also neglect for time being the interaction between $\chi$-fermions. This interaction will renormalize the mass gap, but will not produce any qualitative changes. Then the problem  is equivalent to  the Peierls-Fr\"olich one discussed in the classic soliton theory \cite{braz},\cite{kivelson},\cite{kirova}. Namely, the field theory problem is reduced to the solution of differential equations for the fast Majorana fermion wave functions supplemented by a self-consistency condition:
  \bea
  && E\chi_E = \Big[\ri v_s\hat\tau^3 \frac{\rd}{\rd x} + \Delta(x)\hat\tau^2\Big]\chi_E,\label{Dirac1}\\
  && \Delta(x)/\gamma = \frac{3}{2}\sum_E \mbox{Tr}\Big(\chi_E^*\hat\tau^2\chi_E\Big),\label{Dirac2}
  \eea
  where $\chi_E$ is a two component vector and $\hat\tau^{2,3}$ are the Pauli matrices. It should be emphasized that though $\Delta$-field is static, it is coordinate dependent. 
  
   As was shown in \cite{braz} the latter problem is exactly solvable. All excitations have spectral gaps. For the present purposes it will suffice to consider just a single soliton solution of (\ref{Dirac1},\ref{Dirac2}). The self-consistent solution of (\ref{Dirac1},\ref{Dirac2}) corresponds  to $\Delta(x) = - k_0\tanh[k_0(x-x_0)/v_s]$, where $k_0$ depends on parameters of the model. Dirac equation (\ref{Dirac1}) with such potential has two kinds of solutions. One solution corresponds to $E =0$ and is localized at the soliton center:
   \bea
  \left(
 \begin{array}{c}
 \chi_r\\
 \chi_l
 \end{array} \right)_{k_0 >0} = \left(
 \begin{array}{c}
 1\\
 -1
 \end{array} \right)\exp\Big[\int_0^x \frac{\Delta(y)}{v_s}\rd y\Big]; 
 \left(
 \begin{array}{c}
 \chi_r\\
 \chi_l
 \end{array} \right)_{k_0 < 0} = \left(
 \begin{array}{c}
 1\\
 1
\end{array} \right)\exp\Big[- \int_0^x \frac{\Delta(y)}{v_s}\rd y\Big]. \label{zero}
 \eea 
 The total energy of this solution is finite and originates from the $\Delta^2$-term in (\ref{PF}). Therefore this excitation has a spectral gap (I denote it $M_B$). Its quantum numbers are supplied by Majorana zero modes (\ref{zero}) which also include $\xi^3$. In the limit $u=0$ the soliton does not move; at $u \neq 0$ it acquires dispersion and in that sense it is a slow particle. Further down I will discuss these excitations in more detail. 
 
 Another solution is a scattering state of a (anti)soliton and a massive particle of mass $k_0$: 
 \bea
 && \chi_r(x) = \frac{\re^{\ri kx}}{2\sqrt{L}}\Big[1 + \frac{-kv_s + \ri k_0\tanh[k_0 (x-x_0)/v_s]}{\sqrt{k_0^2 +   (kv_s)^2}}\Big], \nonumber\\
  && \chi_l (x)= \frac{\re^{\ri kx}}{2\sqrt{L}}\Big[1 - \frac{-kv_s + \ri k_0\tanh[k_0 (x-x_0)/v_s]}{\sqrt{k_0^2 +   (kv_s)^2}}\Big],
  \eea
  (these wave functions are normalized) with energy 
  \bea
  E(k) = \sqrt{k_0^2 + (kv_s)^2}.
  \eea
  As I have said, the spectral gap $k_0$ for this excitation is established via self-consistency condition (\ref{Dirac2}). This gap is different from $M_B$, as well as quantum numbers of this massive particle do not coincide with quantum numbers of the slow soliton since its wave function is not localized at $x =x_0$. Namely, the fast particle  carries quantum numbers of the Majorana fermion $\chi^a$ and hence it is S=1 neutral spin exciton. In order to treat quantum corrections to its gap I now reinstate the four-fermion interaction and write down 
  the  effective Hamiltonian for the renormalized fast  modes: 
 \bea
  && H_{fast} =  \label{Int2}\\
  && \frac{\ri v_s}{2}\sum_{a=1}^3(-\chi^a_R\p_x \chi_R^a + \chi_L^a\p_x \chi_L^a) - \ri k_0 \sum_a \chi_R^a\chi_L^a -  2g_{\s,+}\sum_{a>b,a,b=1}^3(\chi_R^a\chi_L^a)(
 \chi_R^b\chi_L^b).
 \eea
 Since $k_0$ is related to the amplitude of $\Delta$,  it follows from (\ref{PF}) that it is related to the vacuum average:  
 \bea
 k_0= -\ri(V+ g\pi v_c - J) \la (\xi_R^1\xi_L^1- \xi_R^2\xi_L^2)\ra \sim \frac{(J - V - \pi gv_c)\ln(\Lambda/M_B)}{4\pi v_c}M_B. \label{mass}
 \eea
 The fermion-fermion interaction in (\ref{Int2}) leads to a substantial renormalization of the bare mass. The latter is determined from  the following equation:
 \bea
 m_s = \frac{k_0}{1 + \frac{g}{2}\ln(\Lambda/m_s)}.\label{mass2}
 \eea
  As it follows from (\ref{mass},  the spin gap $m_s$ becomes smaller with the increase of $v_c$ (that is with the increase of doping). 

 Naturally, in the presence of gapless charge mode $\Phi_c^{(+)}$ the spin exciton is incoherent. Since the staggered spin density operator includes a bosonic exponent of this mode, the exciton is emitted together with a cascade of gapless excitations \cite{rice2}. The resulting composite object is frequently called "magnon-hole-pair bound state", though, strictly speaking this is not a bound state, but a composite. This picture agrees with the numerical results for $\delta =1/8,1/6$ 2-leg ladder obtained in \cite{scalapino1},\cite{scalapino2} and   \cite{gapfunction} for $2\times 24$ and $2\times 12$ ladders respectively. Fig. 1 in \cite{scalapino1} and Fig. 4 in \cite{gapfunction} show
  a sharp peak centered around $\pi(1-\delta)$ which gradually merges into the continuum at larger wave vectors.

  Now I return to the slow modes. As it has been explained above, these are solitons and antisolitons. As was first suggested in \cite{jakiw}, they receive their quantum numbers from the zero modes of Majorana fermions (\ref{zero}) coupled to them (in the present context the reader can also benefit from  the discussion in \cite{CT}).  Such zero modes  ($\xi^3$ mode are included) compose  spinor representations of the O(6) group.  The latter ones  are equivalent to the representations of SU(4) group with Young tableaux consisting of one box (particles) and a column of three boxes (antiparticles). Hence  the solitons carry quantum numbers of electrons (that is transverse momentum and spin) modulo total charge. In other words, the slow excitations are quasiparticles with electric charge stripped into the (quasi) condensate, in short, Bogolyubov quasiparticles.  I denote their spectral gap as $M_B$. As I have mentioned above, I do not see any reason to have different gaps for different transverse momenta as in \cite{scalapino1},\cite{gapfunction}.  Since model (\ref{H0},\ref{Int}) is not Lorentz invariant, a general form the spectrum is not fixed. At small momenta we have 
 \bea
   E_B(p) = M_B + \frac{p^2}{2M^*} + O(p^4),\label{BG}
   \eea
  where $M^* \neq M_B$ due to the lack of Lorentz invariance. To get an idea  of the relative size of the effective mass $M^*$ I notice that  all kinetic energy of a soliton comes from the bare kinetic term in (\ref{PF}). Thus in the leading order in $u/v_s$  the correction to the energy of a moving soliton $\theta(x -vt)$ is
  \bea
  \frac{v^2}{2u} \int \rd x \Big[\frac{\rd\theta}{\rd x}\Big]^2 \sim \frac{v^2}{2u l} \sim M_B(v_s/u)(v/v_s)^2, 
  \eea
  where $l \sim v_s/M_B$ is the solitons's size. As expected, the kink turns out to be a heavy particle: $M^*/M_B \sim 1/uv_s$.  This feature provides consistency to the adiabatic approximation. Since it becomes energetically disadvantageous for zero energy bound states of slow Majoranas $\xi^3$ to "sit" on kinks moving with velocity $v > v_c$,   the region of solitons's stability in momentum space is restricted by  the condition $v < v_c$ or  $|p| < M^* v_c < M_B/v_s$. 
  
 The last remaining problem is whether the underdoped regime may support  d-wave superconducting pairing. As was explained in Sec. IIB, such pairing requires a formation of the proper OP amplitude and this, in turn, requires a proper structure of the vacuum. As follows from (\ref{Int}), the relation between vacua of $\xi$ and $\chi$ fermions  depends on the sign of coupling constants $g_{c,ss} \sim - g_Kv_s  +V+ 3J$, $g_{c.st} \sim (V+  \pi g v_c - J)$ and $g_{\s,-} = (V- J)/2$. When all  these constants are negative 
 the masses of $\chi_a$ and $\xi_1,\xi_3$ fermions have the same sign opposite to the mass of $\xi^2$. As a consequence the average $\la \s_1\s_2\s_3\ra$ forms simultaneously with $\la \Sigma_1 M_2\Sigma_3\ra $ ( or $\la\mu_1\mu_2\mu_3\ra$ with $\la M_1\Sigma_2 M_3\ra$) and hence the entire SCd amplitude (\ref{SCdOP}) freezes (see the discussion in \cite{CT}).  Since $V,J >0$ the conditions to be fulfilled are 
 \bea
 (g_Kv_s -V)/3 > J > V + \pi g v_c. \label{cond}
 \eea

  I conclude this long line of argument with the following comments. 
  
   A qualitative picture of the underdoped regime emerging from these calculations is as follows. First and foremost, when condition (\ref{cond}) for the interchain exchange is fulfilled, the phase has quasi-long-range superconducting order with $d$-wave symmetry. This requires sufficient strength of the exchange interaction which agrees with  the previous expectations. The doping dependence of the spin gap $m_s$ (\ref{mass} and \ref{mass2}) originating from the presence of $v_c$ in (\ref{mass}) is a new result. 
   
   In arrays of ladders SCd order competes with what is variously called Wigner Crystal of Pairs or $4k_F$ Charge Density Wave. The latter phase is probably realized in the telephone number compound \cite{did3}. 
There is a possibility of $m_s$ changing sign before $v_c$ and $v_s$ become comparable, that is within the validity of the adiabatic approximation (this corresponds to a violation of $J > V + \pi gv_c$ condition). This would lead to a transition to $2k_F$ Charge Density Wave phase \cite{CT}.

  Third comment. Usually doping is associated with
incommensurability. However, model (\ref{PF}) looks like commensurate Peierls-Fr\"olich model; the reason for  that is that the only charge 
participating in the game
is the relative one, and the way it couples to the spins reveals
its $Z_2$ nature: it is just $\cos\sqrt{4\pi} \theta$. 
Although the entire system is  doped, and correlations are indeed  incommensurate,
 all this is accounted for by the total charge field $\Phi^{(+)}_c$
which is decoupled from the rest of the system. The spin SU(2) symmetry is,
of course, unbroken, so the spin Majorana mass bilinears can appear only
as a scalar $(\chi^a _R \chi^a _L)$, and a scalar can only couple to
another scalar. The nontrivial conclusion is that as a result of the above the situation for a {\it doped} fermionic ladder 
 is analogous to the {\it commensurate} Peierls-Fr\"olich model with a real 
order parameter $\Delta$. 

Forth comment. By using the results taken from the theory of Peierls-Fr\"olich model I was able to demonstrate that at $v_c << v_s$ the fermion ladder has at least three types of excitations.  They have all been found in the numerical papers \cite{scalapino1},\cite{scalapino2}, \cite{gapfunction}. There are "fast" ($v_s \sim t^2/U$)  spin S=1 magnetic excitons, gapless charge mode $\Phi_c^{(+)}$ and heavy (slow) excitations. The spin  excitations are triplet ones (spin S=1), that is they are qualitatively the same as the excitations of the undoped spin ladder, described by the Hamiltonian similar to (\ref{Int2})  \cite{shelton}. They have a relativistic-like spectrum: 
    \be
    E_s(k) = \sqrt{m_s^2 + (v_sk)^2}. 
    \ee
  As I have said, single S=1 is always emitted in conjunction with the charge gapless modes. The slow excitations also have a spectral gap, but their spectrum is not relativistic (\ref{BG}). By their quantum numbers I identify them as Bogolyubov quasiparticles.     
  It is instructive to compare the described excitation spectrum with the spectrum for the case $v_c \approx v_s$ studied in \cite{fisher},\cite{ludwig}. In the C1S0 phase this  spectrum coincides with  the one for the O(6) symmetric Gross-Neveu (GN) model. It does contain Bogolyubov quasiparticles as kinks and antikinks. In that sense the difference in velocities is not important, as it should be expected for topological excitations. The O(6) GN contains 6-fold degenerate multiplet of vector particles. These ones are local with respect to our Majorana fermions $\xi,\chi$. I have found that at $v_c << v_s$ this multiplet is reduced to 3-fold degenerate multiplet of spin excitons $\chi_a$. Survival of the other three components  including the  famous Cooperons remains an open question.

   \subsection{Single electron Green's functions for single ladder}
   
   Some features of the single particle Green's function whose spectral density is directly measurable by ARPES, can be calculated in the strong coupling regime even if the model (\ref{H0},\ref{Int}) is not integrable. I perform this calculation for the right moving fermions. The decoupling of the total charge mode $\Theta_c^{(+)}$ from the rest of the Hamiltonian translates into factorization of the fermion creation and annihilation operators. Namely, the  creation operator operator for right-moving electron on $j$-th chain  is given by 
   \bea
   \psi^+_{R,\s}(j) = \re^{\ri\sqrt\pi\phi_c^{(+)}}\Big[Z^+_{R,\s,+}\re^{\ri Qx} \pm Z^+_{R,\s,-}\re^{-\ri Qx}\Big] , ~~ (j=1,2), \label{factorization}
   \eea
   where operators $Z^+_{R,\s,p}$ create excitations in the spin and parity sectors, index $p = \pm$ corresponds to  the transverse momenta 0 and $\pi$. The correlation function of the bosonic exponents is known:
   \bea
   \la \re^{\ri\sqrt\pi\phi_c^{(+)}(\tau,x)} \re^{\ri\sqrt\pi\phi_c^{(+)}(0,0)}\ra = (\tau + \ri x/v_c)^{-1/4}\Big[\tau^2 + (x/v_c)^2\Big]^{-(K+1/K-2)/16}\label{Boz}
\eea

The excitation with smallest energy produced by $Z^+$ is a quasiparticle. It can also produce combinations of quasiparticles with magnetic excitons, but these processes require higher energy. Therefore in order to calculate the spectral function close to the low energy threshold $M_B$ we need to determine just one matrix element corresponding to the emission of a quasiparticle. For Lorentz invariant models this can be done rigorously following  the standard procedure explained in  \cite{EssTsv, LukZam}. Although model (\ref{H0},\ref{Int}) is not Lorentz invariant, for energies not very different from the gap $M_B$ when only slow solitons are excited this not be that important. In that case I will just treat (\ref{BG}) as the first terms in the expansion  of $\sqrt{M_B^2 +(vp)^2}$ with $v \sim \sqrt{v_cv_s}$.

Taking these arguments into account I proceed further and assign operators with Lorentz spins.  Recall that operator ${\cal O}_h$ with Lorentz spin $h$ under Lorentz rotation
\bea
vt = (t'v)\cosh\theta + x'\sinh\theta, ~~ x = (t'v)\sinh\theta +x'\cosh\theta
\eea
transforms as 
\bea
{\cal O}_h(x,t) = \re^{h\theta}{\cal O}_h(x',t').
\eea
The right(left)-moving electron is a spinor and hence carries Lorentz spin 1/2 (-1/2), the  chiral bosonic exponent carries Lorentz spin 1/8(-1/8)  and hence the right(left)-moving $Z$-operator must carry Lorentz spin 3/8(-3/8). This means that the matrix element between the vacuum and quasiparticle excitation with energy $\epsilon = M_B\cosh\theta$ and momentum $p = v^{-1}M_B\sinh\theta$ parametrized by rapidity $\theta$ is
\bea
&& \la \theta|Z^+_{R,\s,\pm}(x)|0\ra = {\cal Z}\re^{\pm \ri\pi/4}\re^{3\theta/8 +\ri M_B(x/v)\sinh\theta}, \nonumber\\
&& \la \theta|Z^+_{L,\s,\pm}(x)|0\ra = {\cal Z}\re^{\pm \ri\pi/4}\re^{-3\theta/8+ \ri M_B(x/v)\sinh\theta}. \label{formfactors}
\eea
where ${\cal Z}$ is a normalization constant and $v$ is of the order of $v_c \sim u$. The phase factor $\re^{\pm \ri\pi/4}$ is necessary to maintain the $d$-wave symmetry of the order parameter (see the discussion below). Substituting this expression to the Lehmann expansion for the Green's function and using (\ref{Boz}), I arrive to the following  leading asymptotics of the single electron Green's function: 
   \bea
  &&  \la \psi_{R,\s,\nu}(\tau,x)\psi^+_{R,\s,\nu'}(0,0)\ra \sim \label{Green}\\
   && \delta_{\nu.\nu'}\frac{\re^{\ri \nu Qx}}{(\tau + \ri x/v_c)^{1/4}[\tau^2 + (x/v_c)^2]^{(K + 1/K -2)/16}}\Big(\frac{\tau - \ri x/v}{\tau + \ri x/v}\Big)^{3/8}K_{3/4}\Big(M_B\sqrt{\tau^2 + (x/v)^2}\Big)\nonumber
   \eea
 with $\nu = \pm 1$. Green's function (\ref{Green}) is highly incoherent and is qualitatively similar to the Green's function of the 1/2-filled Hubbard model with different charge and spin velocities calculated in \cite{EssTsv}.  For the sake of simplicity I restrict the calculations for the case $v = v_c$. In that case for $K =1/2$ we get 
 \bea
 && A\Big[\omega,p +\nu Q +k_F, k_{\perp} = \pi(1+\nu)/2\Big] \sim  \label{GreenD0}\\
 && M_B^{-5/16}\frac{(\omega -vp)}{\Big(\omega^2 - (vp)^2 - M_B^2\Big)^{11/16}}F\Big(33/32,9/32,5/16; \frac{M_B^2 + (vp)^2 -\omega^2}{M_B^2}\Big),
 \eea
 where $\nu = \pm 1$. The spectral function vanishes for $\omega^2 < (vp)^2 + M_B^2$.  The above expression is valid below the lesser of two energies: the three-particle threshold $3M_B$ or the threshold for emission of two S=1 excitons $2m_s +M_B$. It is also interesting to note that $Q = t_{\perp}/\sqrt{v_cv_s}$ is non-universal and  can be quite large. Therefore in the presence of interactions the transverse tunneling may shift the spectral weight very substantially from the single chain Fermi points $k_F \pm t_{\perp}/v_F$.
  
\begin{figure}[ht]
\begin{center}
\epsfxsize=0.65\textwidth
\epsfbox{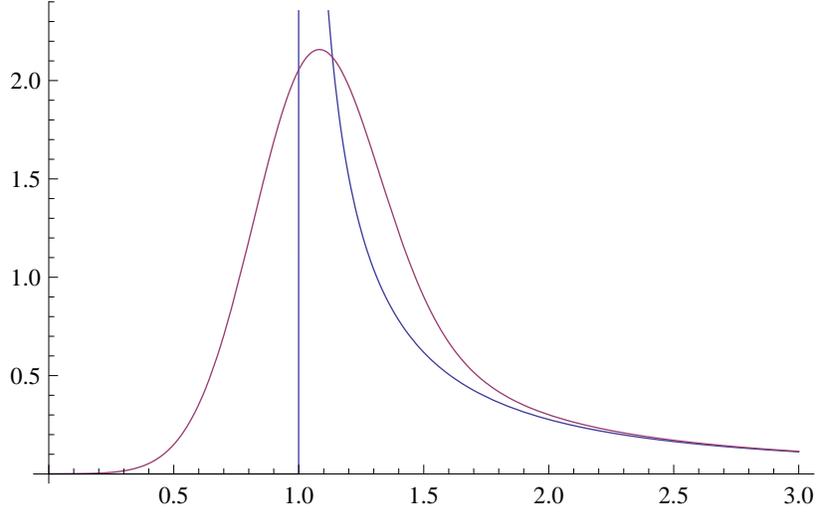}
\end{center}
\caption{The spectral function $A(\omega,\pm Q +k_F) = A(-\omega,\pm Q +k_F)$ (\ref{GreenD0}) as a function of $\omega/M_B >0$ (blue). The sharp maximum at $\omega = \epsilon(k)$  is replaced by a power law singularity. The magenta graph shows the same function convoluted with the Gaussian resolution function $(3/\sqrt\pi M_B) \exp[-9\omega^2/M_B^2]$.}
\label{D0}
\end{figure}

 Two comments are in order. First, in the described energy range the spectral function is particle-hole symmetric which agrees with the early findings \cite{rice2} and disagrees with  the arguments by \cite{pwa}. Particle-hole asymmetry is expected to appear at energies of order of the ultraviolet cut-off and is related to deviations of the bare band dispersion from linearity. Second, although the positions of the peaks in the spectral function depend on the transverse momenta, I have not found any difference in the spectrum for $p_y = 0,\pi$  which is reflected in the identical behavior of $A(k_F+Q)$ and $A(k_F-Q)$. This contradicts the numerical results of \cite{scalapino1}. 
 
 Although the anomalous Green's function vanishes for an infinite ladder, for finite ladder it may be nonzero provided proper boundary conditions are in place. This idea was exploited in \cite{gapfunction},\cite{gapfunc} to extract information about pairing. According to (\ref{factorization}) the anomalous Green's function is 
 \bea
 && F_{ij}(\tau,x) = \la\la \psi_{R,\uparrow,i}(\tau,x_1)\psi_{L,\downarrow,j}(0,x_2)\ra\ra = f(L,\tau,x_1,x_2){\cal F}_{ij}(\tau,x_{12})\nonumber\\
 && f(L,\tau,x_1,x_2) = \la\la \re^{-\ri\sqrt\pi\phi_c{(+)}(\tau,x)}\re^{\ri\sqrt\pi\bar\phi_c^{(+)}(0,0)}\ra\ra,\\
 && {\cal F} = \la\la\Big\{Z_{R,\uparrow,+}(\tau,x)Z_{L,\downarrow,+}(0,0) +[2\delta_{ij}-1]Z_{R,\uparrow,-}(\tau,x)Z_{L,\downarrow,-}(0,0)\Big\}\ra\ra. \label{anom}
 \eea
As I said, the first factor in the last line of (\ref{anom}) vanishes in the infinite system, but is finite in a finite one. It does not contain information about pairing and I do not discuss it. The information about pairing is contained in the second factor which leading asymptotics is determined by the formactors (\ref{formfactors}):
\bea
{\cal F} \sim K_0(M_B\sqrt{\tau^2 + (x/v)^2}) \rightarrow \frac{1}{(\omega+ \ri 0)^2 - (vp)^2 - M_B^2}.
\eea
This expression is valid at frequencies less than $3M_B$. In this region it looks like the standard Gorkov pairing function. The extensive incoherent part, found in \cite{gapfunction},\cite{gapfunc} apparently appears above $3M_B$ when the simple approach described here does not work. 
 

  \subsection{Optimal doping $v_c \sim v_s$}
  
  As it was established already in \cite{fisher}, at $v_c \sim v_s$ the interactions scale together either to strong coupling or to zero.  In the former case the symmetry of the model dynamically increases at low energies and becomes U(1)$\times$O(6). This regime has been thoroughly described in \cite{fisher}, \cite{ludwig}, \cite{fabkon}.  It has been generally assumed that this situation is realized in $t-J$ ladders. However, as it was pointed out in \cite{piotr1},\cite{piotr2}, this is not necessarily the case.  Indeed, 
if one takes as bare couplings (\ref{Bare})  and assumes that $J > V$ (recall that this was one of the conditions for SCd in the underdoped regime) and $vg_K >> V,J ~~ g \sim 1$,  the system will scale to the weak coupling phase C2S2. 
  
 \subsection{Overdoped regime $v_c >> v_s$}
 
  Now the fast fermion mode is  $\xi^3$. For this mode we can repeat the arguments presented in Sec. IIIA. Namely, the fast mode acquires a spectral gap (I denote it $m_3$); its spectrum is
  \be
    E_c(k) = \sqrt{m_3^2 + (v_ck)^2}. 
    \ee 
  Its symmetry is Z$_2$. There are also slow gapful  solitons  whose quantum numbers are the same as for the underdoped regime. Their spectrum is not relativistic. Beyond these statements I can provide no reliable information.  
  
   \begin{figure}
\begin{center}
\epsfxsize=0.45\textwidth
\epsfbox{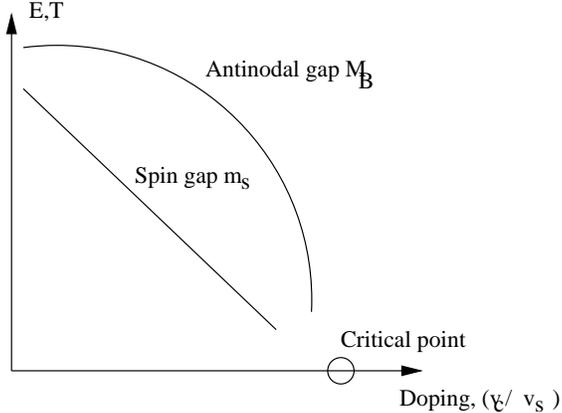}
\end{center}
\caption{The schematic phase diagram of doped 2-leg ladder. } 
\label{rpa}
\end{figure}

 \section{Ladders become stripes.}
 
  The burning question is how properties of a single ladder are going to survive when one assembles ladders together. In this setting doped ladders ("rivers of charge") are separated by undoped ones. Such scenario have been considered many times before in the context of stripe phases  and here the present  approach differs only in details. 
  
   The salient feature of the underdoped 2-leg ladder  is a separation of scales between different excitations. Those ones include  the gapless charge mode,  the fast spin excitons with spin gap $m_s$ and the slow Bogolyubov quasiparticles with gap $M_B$. This separation is likely to survive in the bulk system, at least in some areas of the Brillouin zone. Since charge bosonic exponents enter into expressions for all relevant operators (such as $2k_F$ and $4k_F$ components of spin and charge densities as well as  interladder tunneling operators), transverse propagation of these excitations will be overdamped by emission of gapless phase fluctuations. The coherence will emerge below the 3D superconducting transition when the charge mode condenses (however, see the discussion of the single particle Green's function below).  It is reasonable to assume that the overdamped Bogolyubov quasiparticles of our model correspond to the single electron spectrum in the antinode regions (see a more detailed discussion below). On the other hand, the direct interladder tunneling will give rise to Fermi pockets in the nodal direction, as was found in \cite{KRT}. The precise position of the Fermi pockets at the Brillouin zone diagonals together with their orientation constitutes a difficulty for the present theory which it shares with all models of stripes (see, for instance, \cite{Millis}).
   
   The spin excitons are probably responsible for the famous "hourglass" spectrum observed in the neutron scattering experiments \cite{tranquada}. The calculation presented below is not that different from the one performed in \cite{EssKonTs} where the doped ladder was described by the O(6) Gross-Neveu model. Imagine doped 2-leg ladders running parallel with undoped ones so that a unit cell contains one doped and one undoped ladder. The most singular operator in the spin sector is the $2k_F$ component of magnetization
  \bea
  {\bf S}_- = {\bf n}_-\cos(\sqrt\pi\Phi_c^{(-)})\cos(2k_F x + \sqrt{\pi}\Phi_c^{(+)}) + {\bf n}_+\sin(\sqrt{\pi}\Phi_c^{(-)})\sin(2k_F x + \sqrt{\pi}\Phi_c^{(+)})
 \eea
 Since the sector of $\Phi_c^{(-)}$ is effectively frozen, we can consider all amplitudes to be static:
 \bea
 {\bf S}_- = D{\bf n}_-\cos(2k_F x + \sqrt{\pi}\Phi_c^{(+)}), ~~ D = \la \cos(\sqrt\pi\Phi_c^{(-)})\ra.
 \eea
 So the spin operators in the underdoped regime are almost like the ones  for undoped spin ladder, but slightly "softened" by the gapless charge mode. 
  Then the effective interaction between the staggered magnetization of the undoped chains ${\bf N}$ and the magnetic modes of doped ladders is  
 \bea
 g{\bf N}_k\Big({\bf n}_{-k + \pi\delta} + {\bf n}_{-k - \pi\delta}\Big),
 \eea
 and the RPA spectrum is determined by the equation
 \bea
 \omega^2 - \Delta_0^2 - (Vk)^2 - g^2\cos^2 q\Big[\frac{1}{\omega^2 - m_s^2 - v_s^2(k -\pi\delta)^2} + \frac{1}{\omega^2 - m_s^2 - v_s
 ^2(k +\pi\delta)^2}\Big] =0, 
 \eea
 where $q$ is wave vector transverse to the ladders and $\Delta_0$ is the spectral gap of the undoped ladders. When $m_s < \Delta_0$  this equation yields the hourglass spectrum. 
 
 \subsection{Superconducting fluctuations}

  
   At energies smaller than the quasiparticle gap $M_B$ and spin exciton gap $m_s$ one is left with the gapless charge excitations. In the phase where  SCd order parameter acquires a finite amplitude, the $4k_F$ component of density (\ref{4kF}) also acquires a finite amplitude.  These operators from different chains couple together via Josephson coupling and the Coulomb interaction. This results in the following  effective low energy Hamiltonian for the charge mode (I drop the superscript $(+)$ and subscript $c$):
   \bea
   && H_{charge} = \sum_{{\bf r}}\int \rd x {\cal H}_{\bf r} + U_{Coul} \label{3Dcharge}\\
   && U_{Coul} = \frac{e^2}{2\pi}\sum_{{\bf r} \neq {\bf r}'}\int \rd x \rd x' \frac{\p_x\Phi_{\bf r}(x)\p_{x'}\Phi_{{\bf r}'}(x')}{\sqrt{({\bf r}-{\bf r}')^2+ (x-x')^2}},\label{Coul}\\
   && {\cal H}_n =  \frac{v_c K}{2}\Big[(\p_x\Phi_{\bf r})^2 + (\p_x\Theta_{\bf r})^2\Big] + V_c({\bf r},{\bf r}')\cos\Big[\sqrt{4\pi}(\Phi_{\bf r} - \Phi_{{\bf r}'})\Big] - \nonumber\\
   && V_J({\bf r},{\bf r}')\cos\Big[\sqrt{\pi}(\Theta_{\bf r} - \Theta_{{\bf r}'})- \frac{2e}{c}\int_{{\bf r}'}^{\bf r} \rd{\bf l}{\bf A}\Big]\label{charge1},
 \eea
 where ${\bf A}$ is a vector potential, $V_c$ originates from $4k_F$ component of the Coulomb interaction matrix element and $V_c$ is generated by the interladder Josephson coupling. The long range Coulomb interaction (\ref{Coul}) is often omitted from consideration though it is important to get the right electrodynamics. Model (\ref{3Dcharge}) is directly applicable to the telephone number compound.
 
  It is also instructive to write model (\ref{3Dcharge}) in the Lagrangian form:
  \bea
  && L = \sum_{n=1}^N\int \rd x{\cal L}_n, \\
  && {\cal L}_n = \frac{v_c K}{2}(\p_x\Phi_n)^2 + \ri \p_{\tau}\Theta_n\p_x\Phi_n + \frac{v_c}{2K}(\p_x\Theta_n)^2 + U_{int},
\eea
where interaction term $U_{int}$ contains all cosine terms. At $K =1/2$ the two cosine terms in the interaction have the same scaling dimension and the competition between the superconductivity and Wigner crystal ordering of pairs becomes strong. Recall that $K=1/2$ corresponds to the $U>> t$ limit of the Hubbard model and is a very realistic value in the present context. Then at $V_J > V_c$ the ground state is a superconductor and at $V_J = V_c$ there is a first order phase transition to the Wigner crystal state \cite{carr}. The latter state may emerge even at $V_c < V_J$ if the hopping of pairs between the ladders  is suppressed by a magnetic field. Since field $\Phi$ couples to disorder, In the presence of disorder the pairs localize. It is well known that the suppression of superconductivity in underdoped cuprates reveals the state with weakly insulating properties \cite{bobinger},\cite{bob2}.

In Appendix B I consider exactly solvable model of two coupled ladders. 
 
 \subsection{Single particle Green's function in the striped phase} 
  
   In Section IIIB I studied the Green's function for a single ladder. Now I am going to generalize this calculation for a striped phase. Direct interladder tunneling produces bound states with quantum numbers of electrons. So, one may think that when ladders are coupled together the quasiparticles are reconstituted. However, since the tunneling matrix element is momentum dependent and changes sign throughout the Brillouin zone, the bound states (quasiparticles) are not created at all momenta, but only in those parts of the Brillouin zone where the tunneling process pushes the states below the continuum. It is well known that in the cuprates there are discernible quasiparticle excitations near the nodes and completely incoherent ones at the antinodes \cite{nodes},\cite{STS}. Theoretically emergence of quasiparticle pockets in systems of weakly coupled chains and ladders have been discussed in \cite{KRT}, \cite{EssTsH} and more recently in \cite{subir}.
   
    Below I ignore the direct tunneling which may be a good description for the antinodal region where the  direct tunneling between the ladders does not create quasiparticles.  As far as virtual tunneling processes are concerned, they generate the Josephson coupling between the ladders and are taken into account in effective Hamiltonian (\ref{3Dcharge}).  As for a single ladder case single electron excitation is a Bogolyubov quasiparticle dressed by fluctuations of the gapless charge mode $\Theta_c^{(+)}$. As for a single ladder one can assume that these two types of excitations are decoupled and the electron creation and annihilation operators are  factorized as in (\ref{factorization}) . However, the charge  mode is no longer one-dimensional which makes the calculation slightly nontrivial. 
   
   For the sake of simplicity I will neglect the long distant Coulomb interaction (\ref{Coul}). In the cuprates this may be actually justified in the normal state where the Coulomb interaction is screened by nodal quasiparticles. Then in RPA the charge mode has the phonon-like spectrum 
   \bea
   \Omega^2 = (v_cq_x)^2 + (v/a)\sum_i J_i\sin^2({\bf qe}_i/2),
   \eea
   where ${\bf e}_i$ are the elementary lattice vectors. Above the transition there is a temperature dependent gap, but at the moment I ignore all temperature effects. In the absence of the long range Coulomb interaction model (\ref{charge1}) is Lorentz invariant in ($\tau,x$) space. Therefore one can still assign Lorentz spin to the chiral bosonic exponent  
   \be
   h = \re^{\ri\sqrt\pi\phi_c^{(+)}({\bf r})} \label{h}
   \ee
     and its value is still 1/8. Therefore the matrix element for emission of one bosonic excitation with transverse momentum ${\bf q}_{\perp}$ and rapidity $\theta$ defined as 
   \bea
   \Omega(\theta,{\bf q}_{\perp}) = m({\bf q}_{\perp})\cosh\theta, ~~ q_xv_c = m({\bf q}_{\perp})\sinh\theta, ~~ m^2 = (v/a)\sum_i J_i\sin^2({\bf qe}_i/2)
  \eea 
 is given by 
 \bea
 \la {\bf q}_{\perp},\theta|\re^{\ri\sqrt\pi\phi_c^{(+)}({\bf r})}|0\ra = {\cal Z}^{1/2}({\bf q}_{\perp})\re^{\ri {\bf rq}_{\perp}}\re^{\theta/8},
 \eea
 where ${\cal Z}$ is an unknown coefficient. Since in more than one dimension contribution of multiple emission processes is usually small due to the phase space factors, I will not consider such processes here. The contribution of the single emission to the correlation function of two the chiral exponents positioned on the same ladder is 
 \bea
 \int \rd^D q_{\perp}{\cal Z}({\bf q}_{\perp})\Big(\frac{\tau - \ri x/v_c}{\tau + \ri x/v_c}\Big)^{1/8}K_{1/4}\Big[m({\bf q}_{\perp})\sqrt{\tau^2 + (x/v_c)^2}\Big]
 \eea
 where $D$ is the number of transverse dimensions. Simplifying matters again I take $v_c=v$ and get for the total correlation function 
 \bea
 && G_{RR}(\tau,x; {\bf r} = {\bf r}') = \\
 && \Big(\frac{\tau - \ri x/v}{\tau + \ri x/v}\Big)^{1/2}\int \rd^D q_{\perp}{\cal Z}({\bf q}_{\perp})K_{1/4}\Big[m({\bf q}_{\perp})\sqrt{\tau^2 + (x/v)^2}\Big]K_{3/4}\Big[M_B\sqrt{\tau^2 + (x/v)^2}\Big]\nonumber
 \eea
 Assuming further that $m \sim |q_{\perp}|, {\cal Z}$ = const and $D=1$ for $\omega^2 > M_B^2 + (vp)^2$ we get for the spectral function 
 \bea
 A_{RR}(\omega, p+Q) \sim \frac{\omega - vp}{\omega + vp}\sqrt{\omega^2 - (vp)^2}F\Big(11/8,5/8,1,\frac{M_B^2 + (vp)^2 -\omega^2}{M_B^2}\Big) \label{GreenD1}
 \eea
 
\begin{figure}[ht]
\begin{center}
\epsfxsize=0.65\textwidth
\epsfbox{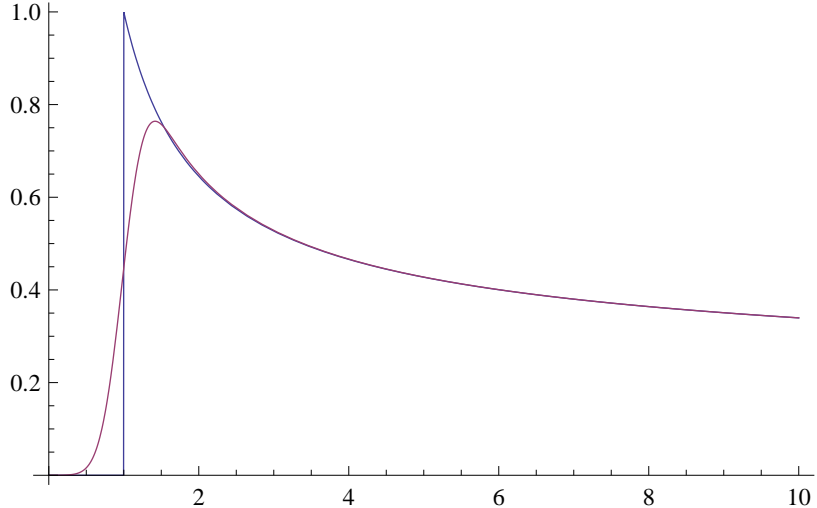}
\end{center}
\caption{The spectral function $A(\omega,Q)$ (\ref{GreenD1}) as a function of $\omega/M_B$ (blue figure). The threshold singularity present for single ladder is removed by the emission of soft phase fluctuations (compare with Fig. (\ref{D0})). The magenta figure represents the spectral function convoluted with the Gaussian resolution function $(3/\sqrt\pi M_B)\exp[-9\omega^2/M_B^2]$.}
\end{figure}

 It may seem counterintuitive  that the increase of dimensionality of the phase fluctuations leads to a stronger suppression of singularity in the Green's function than for a single ladder.  The mechanism  of this increase is related to the factorization of the electron operator already discussed in Section IIIB. This factorization is similar in spirit to the holon-spinon factorization frequently used in gauge field theories of the cuprates (see \cite{LeeWen},\cite{Lee} for a review):
 \be
 \hat\psi_{\s} = \hat h \hat f_{\s},
 \ee
  where operator $\hat h$ carries charge and $\hat f_{\s}$ carries spin. The difference is that holons in that theories are pure bosons and hence can easily condense. The state with condensed holons is Fermi liquid where all incoherence is lost.  There have been always qualitative arguments that the strong gauge field interaction between holons and spinons somehow prevents the former ones from condensation, but these claims have never been substantiated by calculations. In the present case there is no such danger. The role of holon is played by the bosonic exponent (\ref{h}) which has a nonzero Lorentz spin and therefore cannot acquire nonzero average. On the other hand, an object with fractional Lorentz spin becomes very incoherent in more than one dimension which explains the effect.

 \section{Conclusions}
 
  The doped 2-leg ladder demonstrates the most salient features of the bulk cuprate materials. 
  \begin{itemize}
  \item
  When the interchain exchange dominates over the interchange Coulomb repulsion  the phase with one-dimensional analogue of $d$-wave superconducting order parameter is formed. It is understood that for a one-dimensional ladder this means only quasi long range order.
  \item
  When doping increases the system goes through different regimes distinguished by their excitation spectra and behavior of correlation functions. They have many properties similar to underdoped, optimally doped and overdoped regimes of the bulk cuprates. 
  \item
  The excitation spectrum contains both Bogolyubov quasiparticles and collective modes. In the underdoped regime energy scales of all  excitations are well separated. Collective modes include spin-1 neutral excitons and superconducting phase fluctuations. 
  \item
  All excitations are incoherent being overdamped by gapless phase fluctuations. Hence the single particle spectral function does not exhibit a sharp peak, but a broad maximum (pseudogap phenomenon). 
  \item
  The ratio between spectral gaps of quasiparticles and spin excitons is interaction dependent. The spin exciton  gap decreases with doping.
  \item
  The magnitude of all gaps decreases with an increase of doping. There is a possibility that the optimally doped regime where charge and spin velocities are approximately equal is completely gapless. 
 
  \end{itemize}
  
  When ladders are assembled into a quasi-two-dimensional array 
  \begin{itemize}
  \item
  Single particle spectral weight becomes even more incoherent when ladders are assembled in an array and superconducting phase fluctuations cease to be one-dimensional (see the discussion in the end of the previous Section). The long incoherent tail of the spectral function (see Fig. 4) is its an inherent property and not a background.

  \item
  Likewise, a famous hourglass spectrum emerges in the dynamical spin susceptibility of the stripe phase. 
  \item
  Suppression of superconducting coherence by magnetic field drives the system into the insulating state with localized pairs. 
  \end{itemize}
  
   As I have discussed above some of these properties have been detected in the numerical calculations. 
   
  
  \section{Acknowledgements}
  
  I am grateful to A. Chubukov, P. Chudzinski, F. H. L. Essler, T. Giamarchi, P. D. Johnson,  T. M. Rice, J. M. Tranquada and J. Rameau for valueable discussions and encouragement and for D. Poilblanc for bringing me up to date with respect to the numerical work on the problem. My greatest thanks are to A. A. Nersesyan who took trouble to read the manuscript and made extremely valuable comments. This research was supported by the US DOE under contract number DE-AC02-98 CH 10886. 
  
 \appendix 
 
 \section{Bosonization rules for Ising operators}
 
 The Quantum  Ising model 
 \bea
 H = \sum_n\Big(\s^z_n\s^z_{n+1} - h\s^x_n\Big)
 \eea
 is equivalent to the model of a single Majorana fermion. In the continuum limit its spectrum is relativistic $\epsilon(k) = \sqrt{k^2 + m^2}$ where $m = 1-h$. Since two Majoranas constitute one conventional fermion, two Ising models are equivalent to a model of a single  Dirac fermion of mass $m$. The latter model can be bosonized and is equivalent to the  sine Gordon model. Using this chain of equivalencies  one can establish a correspondence between order and disorder parameters of the two Ising models and operators of the Gaussian model. The order parameter operator $\s$ is defined as a continuum limit of $\s^z_n$, its dual operator $\mu$ is a continuum limit of
 \be
 \mu_{n+1/2}^z = \prod_{j=n}^{\infty} \s_j^x
 \ee
 The bosonization rules for Ising model operators are as follows.:
 \bea
 && \cos(\sqrt\pi\Phi) = \mu_1\mu_2, ~~ \sin(\sqrt\pi\Phi) = \s_1\s_2\nonumber\\
 && \cos(\sqrt\pi\Theta) = \mu_1\s_2, ~~ \sin(\sqrt\pi\Theta) = \s_1\mu_2
 \eea
 The above order parameters are sensitive to sign of $m$; $\la\s\ra \neq 0$ for $m>0$ and $\la\mu\ra \neq 0$ for $m<0$. This property is used in the main text of the paper. I refer the reader to \cite{book} for a further discussion of applications of Majorana fermions.

 \section{Phase fluctuations in 4-leg ladder. Exact results.}
 
  To illustrate a competition between Wigner crystallization of pairs and superconductivity, I consider a model of two coupled ladders. 
 I define the modes  
\bea
\Phi_{1,2} = (\Phi_+ \pm \Phi_-)/\sqrt 2
\eea
and the symmetric mode $(+)$ decouples. The Hamiltonian for the asymmetric mode is
\bea
{\cal H}_- = \frac{v_c \tilde K}{2}\Big[(\p_x\Phi_-)^2 + (\p_x\Theta_-)^2\Big] + V_c\cos\Big[\sqrt{4\pi}\Phi_-\Big] - V_J\cos\Big[\sqrt{4\pi}\Theta_-\Big]\label{charge2},
 \eea
where $\tilde K = 2K$ and 
\be
V_c \sim U(4k_F)M_B, ~~ V_J \sim (t^2/M_B)(M_Bm_s)^{3/8}
\ee
 After refermionization I get the sum of two off-critical Ising models:
 \bea
 && {\cal H}_- = \frac{\ri v_c}{2}(-\rho_R\p_x\rho_R + \rho_L\p_x\rho_L - \eta_R\p_x\eta_R + \eta_L\p_x\eta_L) +\nonumber\\
 &&  4\pi v_c (\tilde K -1)\rho_R\rho_L\eta_R\eta_L +  2\ri(V_c + V_J)\rho_R\rho_L + 2\ri(V_c - V_J)\eta_R\eta_L. \label{fin}
 \eea
 Consider $K \approx 1/2$. Neglecting the four-fermion interaction I obtain from (\ref{fin}) that the amplitude of the superconducting order parameter diminishes when $V_J$ approaches $V_c$ from above and vanishes at $V_J \leq V_c$:
 \bea
 \re^{\ri\sqrt{2\pi}\Theta} = \re^{\ri\sqrt\pi\Theta_+}\re^{\ri\sqrt\pi\Theta_-} = \re^{\ri\sqrt\pi\Theta_+}\Big[\mu_1\s_2 + \ri\s_1\mu_2\Big]
 \eea
 The amplitude is 
 \be
 \la \mu_1\s_2\ra \sim (V_J^2-V_c^2)^{1/8}
 \ee
At $V_c > V_J$ one gets the same expression for the amplitude of the $4k_F$ density wave. 

 Magnetic field adds to (\ref{fin}) the term 
 \be
2\ri h(\eta_L\rho_L - \eta_R\rho_R)
\ee
The spectrum becomes
\bea
\omega^2_{1,2} = (v_ck)^2 + h^2 + (V_J^2 + V_c^2) \pm \Big[4(v_ck h)^2 + 4h^2V_J^2 + V_c^2V_J^2\Big]^{1/2}.
\eea
The critical point where $\omega_-(k=0) =0$ is achieved at 
\be
h^2 = (V_J^2-V_c^2)/4.
\ee



\end{document}